\documentstyle[aps,prl,twocolumn,epsf]{revtex}

\begin{document}
\draft
\title{The Nature of the Hall Insulator}
\author{D. Shahar, D.C. Tsui and M. Shayegan}
\address{Department of Electrical Engineering, Princeton University,
 Princeton New Jersey, 08544}
\author{J.E. Cunningham}
\address{ AT\&T Bell Laboratories, Holmdel, New Jersey 07733}
\author{E. Shimshoni}
\address{Department of	Mathematics--Physics, Oranim, Tivon 36006, Israel}
\author{S.L. Sondhi}
\address{Department of Physics, Princeton University,
Princeton New Jersey, 08544}

\date{\today}

\maketitle
\begin{abstract}  
We have conducted an experimental study of the linear transport properties of
the magnetic-field induced insulating phase which terminates the 
quantum Hall (QH) series in two dimensional 
electron systems. We found that a direct and simple relation exists 
between measurements of the longitudinal resistivity, $\rho_{xx}$, 
in this insulating phase and in the neighboring QH phase. In addition,
we find that the
Hall resistivity, $\rho_{xy}$, can be quantized in the 
insulating phase. Our results indicate that a close relation exists 
between the conduction mechanism in the insulator and in the QH liquid.
\end{abstract}

\twocolumn

Quite generally, for two dimensional electron systems (2DES's), the 
quantum Hall effect (QHE) terminates, at high enough magnetic field 
($B$) with a transition to an insulating phase
\cite{Paalanen84,Goldman:wc88,Willett:termination,Jiang90,HWJiang93,Hughes94,Alphenaar:2terminalHI,Shahar:Univ,Dolgopolov:Si92}.
This is true for low 
mobility ($\mu$) samples that exhibit only the integer quantum Hall 
effect (IQHE), for higher mobility samples where the fractional 
quantum Hall effect (FQHE) is resolved and the transition to the 
insulator takes place from the $\nu=1/3$ FQHE liquid state ($\nu$ is 
the Landau level filling fraction), and 
for ultra-high quality samples showing a well-resolved
$\nu=1/5$ FQHE state and the much-discussed reentrant insulating 
phase\cite{Jiang90}.
In the latter, state of the art samples, characterizing and understanding 
the properties of the insulating phase were at the center of an 
intense effort due to the possibility that underlying this insulating 
phase is the elusive $B$--field--induced Wigner solid (WS), expected at low 
$\nu$ for these nearly ideal samples. 
Indeed, most of the experimental results pertinent to this insulator 
are consistent with a WS description\cite{WS:revChoi}.

For lower mobility samples, the transition to the 
insulator occurs at higher $\nu$ (usually $>1/4$) and the formation of 
a WS in not expected. Instead, a novel type of insulator has been 
suggested by Kivelson, Lee and Zhang\cite{KLZ} (KLZ). This insulator, 
named the ``Hall'' insulator, is argued to have a Hall 
resistance whose value is close to the classical value, 
$\rho_{xy} \approx B/ne$, while its
diagonal resistivity, $\rho_{xx}$, diverges as 
$T\rightarrow 0$\cite{Viehweger90,Entin95}.
In contrast to the WS case, a realistic description of 
this suggested new high--$B$ phase is not yet 
available. 

In this letter we present the results of an experimental study of the 
$B$ induced insulator for intermediate and low mobility samples. 
The QHE liquid to insulator transition in such samples, 
which are characterized by a single, well-defined, transition 
to the insulating phase, have been recently studied by Shahar et 
al\cite{Shahar:Univ}.
The main purpose of this letter is to report the 
observation of a new type of symmetry that relates 
this insulating phase to the
adjacent QH liquid phase. When expressed in terms of 
$\rho_{xx}$, this symmetry takes a particularly simple 
form:
 \begin{equation}
 	\rho_{xx}(\nu_{qhe})=1/\rho_{xx}(\nu_{ins})
 	\label{rxxDuality}
 \end{equation}
where, as will be discussed below, $\nu_{ins}$ can be simply obtained from 
$\nu_{qhe}$ by knowing the critical filling factor,
$\nu_{c}$. Eq. \ref{rxxDuality} holds both for transitions from 
the $\nu=1$ IQHE state and, remarkably, also for transitions from 
the $\nu=1/3$ FQHE liquid states. In both cases, the symmetry is 
observed over a rather wide range of $\nu$, and 
$\rho_{xx}(\nu_{qhe})$ can be more than a 100 times smaller than 
$\rho_{xx}(\nu_{ins})$. In addition, this symmetry also holds 
throughout the $T$ range of our study, and for the non-linear 
current dependent regime of transport\cite{Duality:IV}.  
On a phenomenological level, this comprehensive symmetry indicates 
the existence of a direct and simple relation between the conduction 
mechanism in the QHE liquid phases and that of the insulator.

The results presented in this work were obtained from three samples that,
despite being all MBE grown GaAs/AlGaAs heterostructures, 
differ greatly in their parameters. The first two, C70GB and MM051C, 
are low mobility samples (18000 and 11000 cm$^{2}$/Vsec respectively,
at $\sim 1$K)
with very different electron densities of $0.29$ and $2.27\cdot 
10^{11}$ cm$^{-2}$, respectively. In both samples, the disorder level 
is so high that the FQHE can not be resolved, and the IQHE series 
terminates with a direct transition from the $\nu=1$ IQHE state to an 
insulating phase\cite{Shahar:Univ}.
Due to their large density difference, the 
transitions in these samples occur at very different $B$ 
values: $B=1.8$ T for
sample C70GB and $B=14.92$ T for sample MM051C. 
The $\nu_{c}$ values are, however, much closer, 
$\nu_{c}=0.64$ and $0.61$, respectively.

The third sample, M124U2F, has a much higher $\mu$ of $5\cdot 
10^{5}$ cm$^{2}$/Vsec at a density of $0.6\cdot 10^{11}$ cm$^{-2}$. 
Consequently, its low $T$ transport properties are distinctly 
different from that of the other two samples, and include a set 
of FQHE states such as the $\nu=1/3$, $2/5$, $2/3$, $4/3$ etc. In this 
sample, the transition to the insulator occurs directly from the $\nu=1/3$ 
FQHE liquid state, at $B=8.40$ T. We emphasize that the symmetry discussed in this 
letter transcends the large differences in the characteristics of the 
samples in this study.

\begin{figure}
\centerline{\epsfysize=7.5cm
\epsfbox{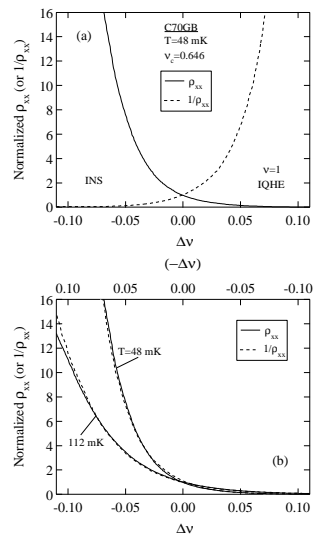}}
\caption{(a) $\rho_{xx}$, normalized to the value of 
$\rho_{xx}$ at $\nu_{c}$, $\rho_{xxc}=23$ k$\Omega$, 
and $1/\rho_{xx}$, calculated from the normalized $\rho_{xx}$ trace,
as a function of 
$\Delta\nu=\nu-\nu_{c}$. (b) A replot of 
the data in (a), but with the $1/\rho_{xx}$ trace plotted against the 
top, $(-\Delta\nu)$ axis. Also included are $\rho_{xx}$ and 
$1/\rho_{xx}$ traces at higher $T$.}
\end{figure}

Our main result is presented in Fig 1, where we plot data obtained 
from sample C70GB. In Fig. 1a, two traces are plotted. The first
(solid curve) is a $\rho_{xx}$ trace obtained at $T=48$ mK, which is 
plotted, rather than versus the usual $B$ 
axis, using $\Delta\nu=\nu-\nu_{c}$ as the abscissa of the graph. 
The $\rho_{xx}$ trace is normalized to its value at $\nu_{c}$, which 
for this sample is $23$ k$\Omega$. The second trace in Fig. 1a is a trace of 
$1/\rho_{xx}$ (dashed line), calculated from the normalized $\rho_{xx}$ 
trace. At the transition point, 
$\rho_{xx}=1/\rho_{xx}=1$. In the IQHE liquid regime, $\rho_{xx}$ 
tends towards zero, and consequently $1/\rho_{xx}$ attains large 
values. In the insulator the opposite trend occurs, with 
$\rho_{xx}$ diverging and $1/\rho_{xx}\rightarrow 0$. 
Plotted in this manner, the two traces appear to be 
reflection--symmetric about the transition point, $\Delta\nu=0$. This 
symmetry can be even more clearly seen in Fig. 1b, where we plot 
the $1/\rho_{xx}$ trace against a reversed, ($-\Delta\nu$), top axis. 
The overlap between the $\rho_{xx}$ and the (reversed) $1/\rho_{xx}$ 
traces is rather good, and extends to the maximum range of our data, 
$\sim 16$ times the transition value, or $\approx 365$ k$\Omega$. The 
extent of this symmetry is even more impressive if we remember that we 
are comparing $\rho_{xx}(\Delta\nu)$ with $1/\rho_{xx}(-\Delta\nu)$, 
where $\rho_{xx}(-\Delta\nu)$, taken in the $\nu=1$ IQHE state is, 
e.g., $1.45$ k$\Omega$ while $\rho_{xx}(\Delta\nu)$ in the insulating 
phase is $365$ k$\Omega$, a ratio of more than $250$ in $\rho_{xx}$.
In Fig. 1b we also include $\rho_{xx}$ and 
$1/\rho_{xx}$ traces taken at $T=112$ mK. Similar symmetry is observed 
for these traces as well, indicating that the symmetry between the QHE and 
the insulator is broader in scope, and is not limited to $\rho_{xx}$ 
at a given $T$, but also 
exists in the $T$ dependence of $\rho_{xx}$. Due 
to lack of space, we 
defer a report of a more complete study of the $T$ dependence of the 
symmetry to a future publication.

The $\rho_{xx}\rightarrow 1/\rho_{xx}$ symmetry,
which extends over a surprisingly wide range of 
$\nu>0.2$, has an interesting experimental consequence. Based only on 
a measurement of $\rho_{xx}$ in, say, the $\nu=1$ IQHE phase it is 
possible to predict, to within experimental accuracy, the value 
of $\rho_{xx}$ at a corresponding point in the insulating regime. The 
corresponding point is determined, for the $\nu=1$ IQHE to insulator 
transition, simply by $\Delta\nu\rightarrow 
-\Delta\nu$, or ``particle--hole'' symmetric points.

\begin{figure}
\centerline{\epsfxsize=5cm
\epsfbox{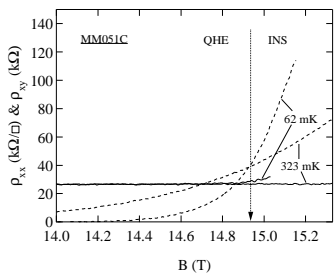}}
\caption{$B$ traces of $\rho_{xx}$ (dashed curves) and $\rho_{xy}$ 
(solid curves) at two temperatures. The arrow indicates the 
transition $B$ (=$14.92$ T) as determined from the crossing point of 
the $\rho_{xx}$ traces.}
\end{figure}

Having shown that a relation exists between the diagonal resistivity 
in the IQHE and the insulator, it is natural to ask if a similar 
mapping can be made with the off--diagonal, Hall component of the 
resistivity tensor, $\rho_{xy}$. This necessitates a careful 
measurement of $\rho_{xy}$ beyond the transition, into the insulating 
phase. This is a notoriously difficult measurement due to the mixing--in 
of a $\rho_{xx}$ component in the $\rho_{xy}$ signal, resulting from 
Hall contacts misalignment and from current path 
non--uniformities\cite{Goldman:1/5Hall,Sajoto:1/5Hall}. While 
the $\rho_{xx}$ component can, in principle, be removed by averaging 
over two $B$ field orientations, in practice the capability to do so 
is usually limited, at low $T$, by inhomogeneity of the current paths. 
Consequently, we could not obtain reliable $\rho_{xy}$ data in the 
insulating regime for any of our samples at the lower end of our $T$ 
range ($T<100$ mK). At higher $T$, and for sample MM051C, 
which is of very high density and 
therefore tends to be less prone to inhomogeneities, we were able to 
follow the evolution of $\rho_{xy}$ into the insulating phase. In Fig. 2 
we present $B$ traces of $\rho_{xy}$ (solid lines) near the critical $B$, 
$B_{c}=14.92$ T (arrow in Fig 2). $B_{c}$ is determined by the 
crossing point of the $\rho_{xx}$ 
traces (dashed lines) taken at different $T$'s. For this sample, only 
minor ($<0.5\%$) variations in $B_{c}$ exist between measurements 
using different contact configurations, attesting to an exceptionally 
high degree of homogeneity. 
The main observation that can be made from these $\rho_{xy}$ traces
is that at $T=323$ mK, $\rho_{xy}$ remains constant and {\em 
quantized}, $\rho_{xy}=h/e^{2}$, for a 
finite $B$ range into the insulating phase.
On the other hand, at $62$ mK, 
the measured $\rho_{xy}$ begins to deviate from its 
quantized value at $B<B_{c}$. At present, we are unable to determine 
whether this deviation arises from the intrinsic behavior of 
$\rho_{xy}$, or from an extrinsic source, such as the 
rapidly--increasing $\rho_{xx}$ contribution to the measured 
$\rho_{xy}$, or a development of current non-uniformities at lower $T$.
Further studies are required to 
establish the range of $B$ and $T$ where $\rho_{xy}$ remains quantized in the 
insulator, and to determine whether its asymptotic behavior is that 
of a ``Hall'' insulator, as predicted by KLZ. Interestingly, a 
constant $\rho_{xy}$ is a consequence of the `semi--circle' 
conductivity law recently suggested theoretically 
by Ruzin and his collaborators\cite{Ruzin:semicircle}.  

\begin{figure}
\centerline{\epsfxsize=5cm
\epsfbox{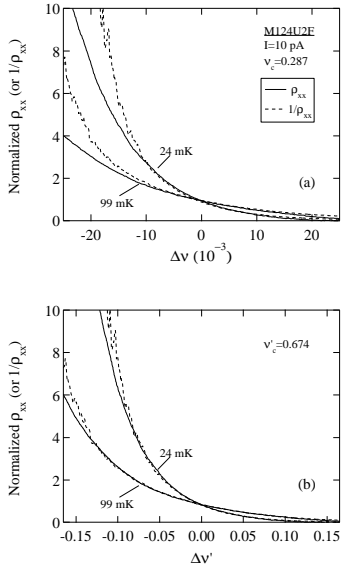}}
\caption{(a) Normalized $\rho_{xx}$ vs.
$\Delta\nu=\nu-\nu_{c}$ and $1/\rho_{xx}$ vs. $-\Delta\nu$ 
for the $\nu=1/3$ to insulator transition. 
For this sample, $\rho_{xxc}=21$ k$\Omega$.
(b) Same data as in (a), plotted using the filling factor of 
composite fermions, $\nu'$ (see text). Note the improved 
$\rho_{xx}=1/\rho_{xx}$ symmetry.
}
\end{figure}

It is now rather well--accepted that strong similarities exist 
between transitions in the IQHE and FQHE regimes
\cite{Shahar:Univ,LWEngel90,Wong95}.
It is therefore 
interesting to look for the symmetry near transitions from FQHE 
states to the insulator. 
In Fig. 3a we plot two $\rho_{xx}$ traces (solid curves)
vs. $\Delta\nu$, obtained near the 
$\nu=1/3$ to insulator transition, at $T=24$ and $99$ mK. Similarly to Fig. 
1, we also plot $1/\rho_{xx}$ traces (dashed curves)
vs. a reversed $\Delta\nu$ axis. Although 
very close to the transition point we again obtain the 
$\rho_{xx}=1/\rho_{xx}$ symmetry, the range in terms of $\rho_{xx}$ 
appears to be  
much more limited. We now recall that, in the KLZ description of the 
transitions in the FQHE regime, the electronic system is mapped,
using a Chern--Simons transformation, onto a superconductor--insulator
transitions of an equivalent composite boson (CB) system\cite{ZHK89}.
A similar transformation was used by Jain\cite{Jain89} to map
the $\nu=1/3$ FQHE state onto a $\nu=1$ IQHE state of composite 
fermions (CF's). We will next show that, by mapping our $\nu=1/3$ to 
insulator transition onto a $\nu=1$ to insulator transition of CF, 
the $\rho_{xx}=1/\rho_{xx}$ symmetry greatly improves, and the 
corresponding $\nu$'s are again related by particle--hole symmetry.
Alternatively, it is possible to use the CB's mapping,
and the corresponding $\nu$'s are then related by 
duality--symmetry of the CB's\cite{LeeFisher:AnyFQHE}.

Using the 
CF mapping, we replot in Fig. 3b the same data as in Fig. 3a, 
only with the fillings of CF, $\nu'=\frac{\nu}{1-2\nu}$, as the abscissa. 
A much--improved symmetry is obtained at both $T$'s, 
extending to $\sim 6$ times the transition value at $24$ mK and $\sim 
4$ times at $99$ mK, before appreciable systematic deviations are 
observed. 
In view of the data in Fig. 3, it is reasonable to
assume that for the $\nu=1/3$ to insulator transition, the 
symmetry is that of the underlying composite particles, rather than 
that of the electrons. 

\begin{figure}
\centerline{\epsfxsize=5cm
\epsfbox{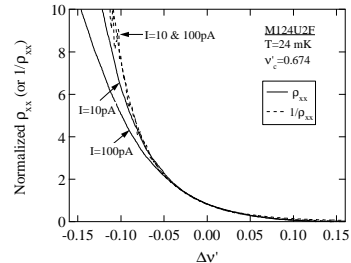}}
\caption{Normalized $\rho_{xx}$ vs.
$\Delta\nu'$ and $1/\rho_{xx}$ vs. $-\Delta\nu'$, as in Fig. 3b,
but for two currents, at $T=24$ mK.}
\end{figure}

Although we can not give a definite answer to the question of the 
range of validity of the $\rho_{xx}\rightarrow 1/\rho_{xx}$ symmetry, 
an illuminating experimental point can be made by considering the 
current dependence of the $\rho_{xx}$ traces. In Fig. 4, we again plot 
$\rho_{xx}$ and $1/\rho_{xx}$ traces, similar to those in Fig. 3, 
taken at $T=24$ mK, but measured at two different currents. Clearly, 
the systematic deviation of the $\rho_{xx}$ traces (solid lines) from 
the $1/\rho_{xx}$ traces (dashed lines) is much larger, and starts 
closer to $\nu'_{c}$, for the $100$ pA trace. 
On the other hand, the $1/\rho_{xx}$ traces 
at both currents appear to overlap (for the $1/\rho_{xx}>1$ range),
indicating that the $\rho_{xx}$ in 
the FQHE regime is independent of current in 
the range shown. It therefore appears that by lowering the measuring 
current the range of applicability of the symmetry is significantly 
increased, even at these very low currents. 
A natural conclusion would be that the symmetry holds only in 
the strictly ohmic regime of transport. We wish to emphasize here, that this 
is not the case, and that the symmetry is maintained much beyond the 
ohmic regime. 
However, since for higher currents outside the 
ohmic regime, $\rho_{xx}$ is ill--defined, a generalized symmetry
should be invoked. This can be done by considering the 
$\rho_{xx}\rightarrow 1/\rho_{xx}$ symmetry as a symmetry under the 
exchange of the current and voltage coordinates. In a separate 
publication, we demonstrate that this generalized symmetry 
indeed holds for a broad range of currents and voltages\cite{Duality:IV}. 

From the discussion in the previous paragraph, we conclude that it is 
essential for a quantitative comparison of $\rho_{xx}$ on both sides of 
the transition that the measurement current will be kept to a minimum 
to prevent non--ohmic effects which, for a constant--current 
measurement, arise predominantly in the insulator. 
With the exception of Fig. 4, we use an 
AC current of $10^{-11}$ A and monitor the 
voltage by a lock--in phase sensitive technique. We also use very slow $B$ 
sweep rates ($>15$ min/T) to maintain $T$ stability at dilution 
temperatures $\geq 20$ mK. We found that using these slow sweep 
rates, our Oxford 200 TLM refrigerator maintained $T$ stability of 
better than $2\%$ throughout our $T$ range.

The results presented in this letter pose a rather strict constraint 
on any theoretical model aiming at describing the QHE to insulator 
transitions. We wish to outline here 
some of the features required from such a model. First, the model has to 
be particle--hole symmetric ($\Delta\nu \leftrightarrow -\Delta\nu$), 
as is evident from the $\nu$ values of the corresponding states in 
the insulator and the QHE states. For the $\nu=1/3$ to insulator 
transition, the symmetry should be that of the composite particles. 
Second, the model has to 
allow for the symmetry under $\rho_{xx}\rightarrow 1/\rho_{xx}$ exchange, 
i.e. an exchange of the measured currents and voltages. This symmetry 
can be conveniently described as symmetry under duality transformation,
and is further discussed in a separate paper\cite{ShimshoniDuality}.

While our experiment does not directly identify 
the mechanism for ohmic transport in the Hall insulator, it does
show that this mechanism is very
closely related to that governing the conduction process in the IQHE 
and FQHE regimes, over a surprisingly wide range of $\nu$.
It would be interesting to look for a similar symmetry near
other transitions, such as the QHE plateau to plateau 
transitions, transitions to an insulating phase
in double--layer QHE systems, and perhaps even the 
superconductor--insulator transitions in thin metal films and in 
Josephson junction arrays. 

\noindent
We thank S. A. Kivelson for instructive discussions.
This work has been supported by the NSF, the Beckman Institute (ES)
and the A. P. Sloan Foundation (SLS).


\begin{references}

\bibitem{Paalanen84} M.A. Paalanen,	D.C. Tsui, A.C.	Gossard	and	J.C.M. 
Hwang, Solid State Comm. {\bf 50} 841 (1984).
\bibitem{Goldman:wc88} V. J. Goldman, M. Shayegan and D. C. Tsui,
Phys. Rev. Lett. {\bf 61}, 881 (1988).
\bibitem{Willett:termination}
R. Willett {\it et~al.}, Physical Review B {\bf 38},  7881  (1988).
\bibitem{Jiang90} H.W. Jiang {\it et al.}, Phys. Rev. Lett. 
{\bf 65}, 633 (1990).
\bibitem{HWJiang93}
H. W. Jiang and C. E. Johnson and K. L. Wang and S. T. Hannahs, 
Phys. Rev. Lett. {\bf 71},  1439  (1993).
\bibitem{Hughes94} R.J.F. Hughes {\it et al.}, J. Phys. 
Condens. Matter {\bf 6}, 4763 (1994).
\bibitem{Alphenaar:2terminalHI} B. W. Alphenaar and D. A. Williams,
Phys. Rev.	B {\bf	50}, 5795 (1994). 
\bibitem{Shahar:Univ} D. Shahar, D. C. Tsui, M. Shayegan, R. N. Bhatt and
J. E. Cunningham, Phys. Rev. Lett. {\bf 74}, 4511 (1995).
\bibitem{Dolgopolov:Si92} V. T. Dolgopolov, G. V. Kravchenko, A. A. 
Shashkin and S. V. Kravchenko, Phys. Rev. B {\bf 46}, 13303 
(1992).
\bibitem{WS:revChoi} For a recent review see 
{\em Physics of the electron solid}, edited by S. T. Chui,  
(International Press, Boston, 1994).
\bibitem{KLZ} S. A. Kivelson, D. H. Lee, and S. C. Zhang, Phys. Rev. B
{\bf 46}, 2223 (1992).
\bibitem{Viehweger90} O. Viehweger and K. B. Efetov, J. Phys. Cond. 
Mat. {\bf 2}, 7049 (1990).
\bibitem{Entin95} See also O. Entin-Wohlman, A.G. Aronov, Y. Levinson and Y. 
Imry, Phys. Rev. Lett. {\bf 75}, 4094 (1995); and references therein.
\bibitem{Duality:IV} D. Shahar, D. C. Tsui, 
M. Shayegan, E. Shimshoni and S. L. Sondhi (unpublished).
\bibitem{Goldman:1/5Hall} V. J. Goldman, J. K. Wang, 
Bo Su and M. Shayegan, Phys. Rev.	Lett. {\bf 70},	647 (1993).
\bibitem{Sajoto:1/5Hall} T. Sajoto, Y. P. Li, L. W. Engel, D. C. Tsui 
and M. Shayegan, Phys. Rev.	Lett. {\bf 70},	2321 (1993).
\bibitem{Ruzin:semicircle} A. M. Dykhne	and	I. M. Ruzin, Phys. Rev.	B {\bf 50},
2369	(1994);	I. M. Ruzin	and	S. Feng, Phys. Rev.	Lett. {\bf 74},	154
(1995).
\bibitem{LWEngel90} L.W. Engel, H.P. Wei, D.C. Tsui, and M. Shayegan, 
Surf. Sci. {\bf 229}, 13 (1990).
\bibitem{Wong95} L. W. Wong, H. W. Jiang, N. Trivedi and E. Palm,
Phys. Rev. B {\bf 51}, 18033 (1995).
\bibitem{ZHK89} S.C. Zhang, T. H. Hansson and S. A. Kivelson, Phys. Rev. Lett.
{\bf 62}, 82 (1989).
\bibitem{Jain89} J.K. Jain, Phys. Rev. Lett. {\bf 63}, 199 (1989); 
Phys. Rev. B {\bf 40}, 8079 (1989).
\bibitem{LeeFisher:AnyFQHE} D.-H. Lee and M. P. A. Fisher, Phys. Rev. Lett.
{\bf 63}, 903 (1989); For
discussions on	duality see M. P. A. Fisher, Phys.	Rev. Lett. {\bf	65},
923 (1990) and C. A. L\"{u}tken and G. G. Ross, Phys. Rev. B {\bf 48},
2500 (1993).
\bibitem{ShimshoniDuality} E. Shimshoni, S.	L. Sondhi and D. Shahar	(in	preparation).
\end{references}
\end{document}